\documentclass[fleqn,usenatbib]{mnras}

\usepackage[T1]{fontenc}

\DeclareRobustCommand{\VAN}[3]{#2}
\let\VANthebibliography\thebibliography
\def\thebibliography{\DeclareRobustCommand{\VAN}[3]{##3}\VANthebibliography}

\usepackage{graphicx}	% Including figure files
\usepackage{amsmath}	% Advanced maths commands
\usepackage[normalem]{ulem}
\usepackage{newtxtext,newtxmath}
\usepackage{color}

\def \feix {Fe\,{\sc ix}}
\def \fexv {Fe\,{\sc xv}}
\def \fexix {Fe\,{\sc xix}}

\title[MUSE synthetic spectra]{MUSE observations of small-scale heating events}

\author[C. A. Breu et al.]{
C. A. Breu,$^{1}$\thanks{E-mail: cab42@st-andrews.ac.uk (University of St. Andrews)}, I. De Moortel,$^{1,2}$, P. Testa$^{3}$
\\
$^{1}$School of Mathematics and Statistics, University of St Andrews, St Andrews, Fife KY16 9SS, UK\\
$^{2}$Rosseland Centre for Solar Physics, University of Oslo, PO Box 1029 Blindern, NO-0315 Oslo, Norway\\
$^{3}$Harvard-Smithsonian Center for Astrophysics, 60 Garden St, Cambridge, MA 02193, USA\\}
\date{Accepted XXX. Received YYY; in original form ZZZ}

\pubyear{2024}

\begin{document}
\label{firstpage}
\pagerange{\pageref{firstpage}--\pageref{lastpage}}
\maketitle

\begin{abstract}
Constraining the processes that drive coronal heating from observations is a difficult task due to the complexity of the solar atmosphere.
 As upcoming missions such as MUSE will provide coronal observations with unprecedented spatial and temporal resolution,
 numerical simulations are becoming increasingly realistic.
 Despite the availability of synthetic observations from numerical models,  line-of-sight effects and the complexity of the magnetic topology in a realistic setup still complicate the prediction of signatures for specific heating processes.
 3D MHD simulations have shown that a significant part of the Poynting flux injected into the solar atmosphere is carried by small-scale motions, such as vortices driven by rotational flows inside intergranular lanes. MHD waves excited by these vortices have been suggested to play an important role in the energy transfer between different atmospheric layers. 
Using synthetic spectroscopic data generated from a coronal loop model incorporating realistic driving by magnetoconvection, we study whether signatures of energy transport by vortices and eventual dissipation can be identified with future missions such as MUSE. 
\end{abstract}

\begin{keywords}
Sun: corona -- Sun: magnetic field -- (magnetohydrodynamics) MHD -- Sun: UV radiation
\end{keywords}

%%%%%%%%%%%%%%%%%%%%%%%%%%%%%%%%%%%%%%%%%%%%%%%%%%

%%%%%%%%%%%%%%%%% BODY OF PAPER %%%%%%%%%%%%%%%%%%

\section{Introduction}

Spectroscopic measurements allow us to study the physical processes responsible for coronal heating. The upcoming Multi-Slit Solar Explorer (MUSE) mission is expected to provide spectroscopic data with unprecedented spatial and temporal resolution as well as spatial coverage \citep{2020ApJ...888....3D, 2022ApJ...926...52D}. The MUSE spectrograph will observe in three extreme ultraviolet wavelength channels centered around 171 \AA, 284 \AA\ and 108 \AA, including strong unblended lines of \feix, \fexv, and \fexix, respectively, formed around temperatures of 0.8 MK up to 12 MK. This large span of plasma temperatures allows for the observation of a wide range of phenomena in different layers of the atmosphere, from the transition region to flare plasma. Due to its multi-slit nature (the MUSE spectrograph will have 35 slits), MUSE has a large spatial coverage, providing simultaneous spectra of entire structures such as coronal loops, while having a high spatial resolution of $0\farcs 4\times 0\farcs 167$ \citep{2022ApJ...926...52D}, with the slit width $0\farcs 4$ corresponding to 291 km and the resolution along the slit to 121.409 km on the solar surface. %This resolution is achieved along the central 140\arcsec of the slit, covering e region of roughly 100 Mm. (EUVST)
In contrast, existing coronal spectrometers such as Hinode/EIS have a much lower resolution of over 2\farcs.\\
A high spatial and temporal resolution is vital since heating events in the corona are thought to take place on small spatial scales and short timescales.
In recent years, simulations and observations have shown that small-scale motions in the intergranular lanes, especially vortex motions driven by magnetoconvection may play an important part in the transfer of energy and mass to the corona, e.g. \citep{2012Natur.486..505W,2020arXiv200413996Y,2021A&A...645A...3Y,2023ApJ...949....8K}. 
These flows are difficult to detect due to their small spatial extent, down to 0.58 Mm in the chromosphere \citep{2019NatCo..10.3504L} and below 100 km in the photosphere. 
Nevertheless, vortices have been detected in  the photosphere and in the chromosphere \citep{1988A&A...198..322B, 2008ApJ...687L.131B, 2010ApJ...723L.139B,2009A&A...507L...9W}, where some of the detections were associated with brightenings in the low corona indicative of heating \citep{2012Natur.486..505W}. Due to the spatial coincidence of detections across multiple atmospheric layers, it is hypothesized that swirls consist of a coherent, rotating or twisted magnetic field structure connecting several atmospheric layers.
Simulation results indicate that chromospheric swirls with sizes of several Megameters could be made up of smaller swirls of the size of a few kilometers \citep{2021A&A...645A...3Y}, indicating a turbulent cascade.
It is unclear how many of these structures penetrate the transition region and continue into higher layers of the atmosphere, but it has been suggested that they could launch torsional Alfv\'{e}n perturbations into the corona \citep{2013ApJ...776L...4S,2021A&A...649A.121B}.
In order to observe wave signatures, observations at high cadence would be required in addition to very high spatial resolution, due to the low densities and high Alfv\'en speed in the corona.
In the paper, we will investigate whether MUSE would be able to detect atmospheric swirls and follow a propagating twist along a coronal loop.

\section{Methods}

\subsection{Simulation Setup}

We model a coronal loop as a straightened-out magnetic flux tube in a Cartesian box with dimensions $6\times 6\times 57$ Mm and a spatial resolution of $\Delta =60$ km. 
The 3D resistive MHD simulations are solved using the MURaM code \citep{2005A&A...429..335V} with the coronal extension \citep{2017ApJ...834...10R}.
The effects of gravitational stratification, field-aligned
heat conduction, optically thick gray radiative transfer in the
photosphere and chromosphere and optically thin losses in the
corona are taken into account. As an initial condition for the magnetic field we use a uniform vertical field with a field strength of 60 G. In the photosphere, the magnetic field is then concentrated into the intergranular lanes by convection.
The setup is described in more detail in \citet{2022A&A...658A..45B}.\\
Using a straightened-loop model in combination with a realistic treatment of the magnetoconvection at the loop footpoints, we found that small-scale torsional motions in the intergranular lanes play a non-negligible role in the energy transport in coronal loops and the model reproduced observed strand widths of coronal loops \citep{2013ApJ...772L..19B,2020ApJ...902...90W}.

\subsection{Differential Emission Measure Calculation}

For optically thin conditions, the differential emission measure (DEM) quantifies the contribution to the emission by plasma within a specific temperature interval.
The energy flux F in a certain emission line at a specific location is given by a height integration of the emissivity along the line-of-sight (LOS) \citep{2006ApJ...638.1086P}:
\begin{equation}
    F = \int G(T,n_{e})n_{e}^{2}ds,
\end{equation}
where $n_{e}$ is the electron density, $G(T,n_{e})$ the contribution function of the respective emission line and $ds$ the line element along the LOS.\\
Since the integral is not dependent on the ordering of volume elements along the LOS, this can be replaced by an integration over the temperature \citep{1976A&A....49..239C}:
\begin{equation}
    F = \int G(T,n_{e})\rm{DEM}\; dT
\end{equation}
with the differential emission measure 
\begin{equation}
    \rm{DEM} = n_{e}^{2}\frac{ds}{dT}.
\end{equation}
Due to the Doppler effect, emission line profiles produced by a moving plasma parcel will be shifted to longer (if moving away from the observed) or shorter (if moving toward the observer) wavelengths, so the energy flux due to emitted photons of a given wavelength depends on the plasma velocity in addition to the temperature. In order to obtain synthetic spectra, we need to obtain the distribution of emission as a function of plasma temperature and the velocity component along the LOS. This quantity is termed the velocity differential emission measure (VDEM) \citep{1995ApJ...447..915N}:
\begin{align}
    \rm{VDEM} &= n_{e}^{2}\frac{ds}{dTdv},\\
    F &= \int\int G(T,n_{e})\rm{VDEM}\; dT\;dv.
\end{align}
In practice, we divide the temperature and velocity range present in the simulation into intervals and determine the amount of plasma present in each temperature and velocity bin.\\

From the simulation cubes, we obtain density, temperature and the line-of-sight velocity.
%The density is converted to electron number density by dividing by the proton mass $m_{p}=1.6726219\times 10^{-24}\; \rm{g}$. 
The electron density is computed using the tabulated equation of state (EOS).

%The units are as follows: $\rm{cm^{-3}}$ for the electron density, K for the temperature and $\rm{km\ s^{-1}}$ for the LOS-velocity and $\rm{cm}$ for ds (here the grid spacing).

%Here we choose 51 bins for both temperature and Doppler velocity and then calculate the contribution to the emission from plasma in a certain temperature and velocity bin.
We choose a linear spacing for the LOS-velocity and logarithmic spacing for the temperature.
The width of the temperature bins in logarithmic space is 0.1, which is a typical bin width for a DEM. The width of the velocity bins is $5\; \rm{km\; s^{-1}}$. 

\subsection{Synthetic spectra}

With the obtained VDEM and the MUSE response function, we can then calculate the synthetic spectrum for a specific spectral line: 
\begin{equation}
    I_{\lambda}(\mathbf{x},\lambda)= \rm{VDEM}_{tvij}r^{vtm}ds,
\end{equation}
where $r$ is the instrument response function for MUSE providing the detector response for all 1024 spectral bins for all three wavelength channels per unit emission measure (in $10^{-27}\; \rm{cm^{-5}}$ ), t and v are the indices along the temperature and velocity axes, i and j are the spatial indices of each pixel and m is the index of the spectral bin. \\
The response function has been calculated using CHIANTI 10.0 assuming coronal abundances and includes instrumental line broadening and thermal line broadening \citep{2022ApJ...926...52D}. The synthetic spectra are in units of $\rm{[ph/s/pix]}$ with the MUSE spectral pixel of size of $0.\arcsec 4\times 1.\arcsec 67$.

\subsection{Spectral Profile Moments}

After synthesizing the spectral line profiles, we calculate the profile moments to obtain intensity, Doppler shift and line width:
\begin{align}
    I_{0}(\mathbf{x})&=\int I_{\lambda}(\mathbf{x}, \lambda)d\lambda,\\
    I_{1}(\mathbf{x})&=\frac{1}{I_{0}(\mathbf{x})}\int \lambda I_{\lambda}(\mathbf{x}, \lambda)d\lambda,\\
    I_{2}(\mathbf{x})&=\sqrt{\frac{\int (\lambda -I_{1}(\mathbf{x}))I_{\lambda}(\mathbf{x}, \lambda)}{I_{0}(\mathbf{x})}}d\lambda.
\end{align}
This gives the intensity in units of ph/s/pix and the Doppler shift and line width in \AA. 
From the second profile moment we calculate the exponential line width $\sigma_{1/e}(\mathbf{x})=\sqrt{2}I_{2}(\mathbf{x})$.
The wavelength is in units of \AA.
To convert from wavelengths to Doppler velocities, we use $v_{\rm{source}}=c\left(\frac{\lambda'}{\lambda}-1\right)$, where $v_{\rm{source}}$ is the Doppler velocity of the emitting plasma, $c$ is the speed of light, $\lambda$ is the unshifted wavelength and $\lambda'$ is the observed wavelength.\\

Since the MUSE resolution differs from the grid resolution of the simulation, we resample our obtained spectra to the MUSE plate scale. 
Following \citet{2022ApJ...926...52D}, we resample the spectra to a pixel size of $0\farcs 167\times 0\farcs 167$ since the slit could always be reorientated to obtain the maximum possible resolution. For the raster scan in Fig. \ref{fig:raster_scan}, however, we use a resolution of (0.4, 0.167)\arcsec. Since the grid spacing of the simulation is $\Delta=60\; \rm{km}$, we approximate the MUSE resolution by averaging the spectra from a $2\times 2$-pixel wide square.

\subsection{Data analysis}
\label{sect:datanalysis}

We compute MUSE synthetic spectra for two different observing modes, a raster scan and a sit-and-stare observation. In the first case, the slits are moved over the observed structure, whereas the location of the slits is kept constant in the latter case so that the time evolution of the spectra can be studied in a fixed location. For the raster scan we assume an exposure time of 1 s, so each column of pixels parallel to the slit has a time difference of 1 s.\\
Small-scale torsional motion has been suggested to cause an increase in line broadening \citep{2015ApJ...799L..12D}.
A swirl seen at the limb could manifest either as adjacent red- and blueshifted features or as region with broadened emission lines if several structures are overlapping along the LOS or have sizes below the instrument resolution limits.
We compute MUSE spectra for 14 different slits with locations marked by dashed black lines in Fig. \ref{fig:raster_scan}.
We systematically check for events showing high line broadening in time-distance diagrams for the different MUSE slits.\\
To identify an event, we compute the average line broadening over the time series for each slit. We define an event as line broadening exceeding the average broadening by five times the standard deviation. This threshold was chosen to filter out only the strongest events associated with clusters of swirls or strong shear flows.
At the location of the peak of the event, we check for increased Doppler shifts and increased emission in different lines.

\section{Results}

\subsection{Detected Events}
\label{section:events}

\begin{figure*}
	\includegraphics[width=19cm]{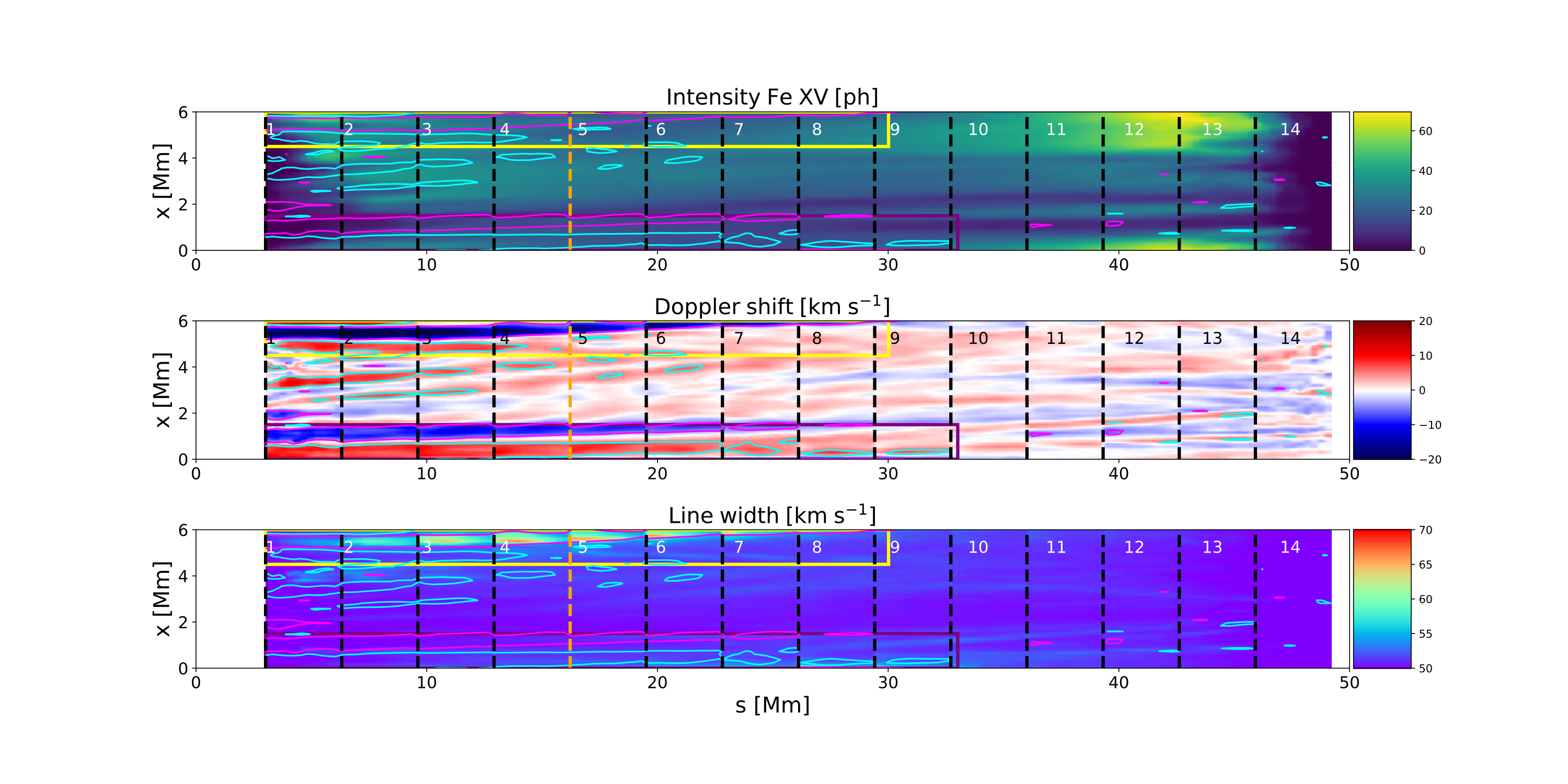}
    \caption{MUSE synthetic raster scan of the coronal loop model for the \fexv\ 284 \AA\ line. From top to bottom: Intensity, Doppler shift and line width. We assume a pixel size of (0.4, 0.167)\arcsec and an exposure time of 1 s. The loop was scanned from left to right. The dashed black lines illustrate the location of the MUSE slits at the start of the exposure. Slit 5 is chosen for closer examination and is marked in orange. The raster scan covers the strongest line broadening event in the time series. The yellow and purple rectangles mark the location of strong adjacent red- and blueshifts. The light blue and pink contours outline Doppler shifts stronger than $5\; \rm{km\; s^{-1}}$.}
    \label{fig:raster_scan}
\end{figure*}

\begin{figure*}
	\resizebox{\hsize}{!}{\includegraphics{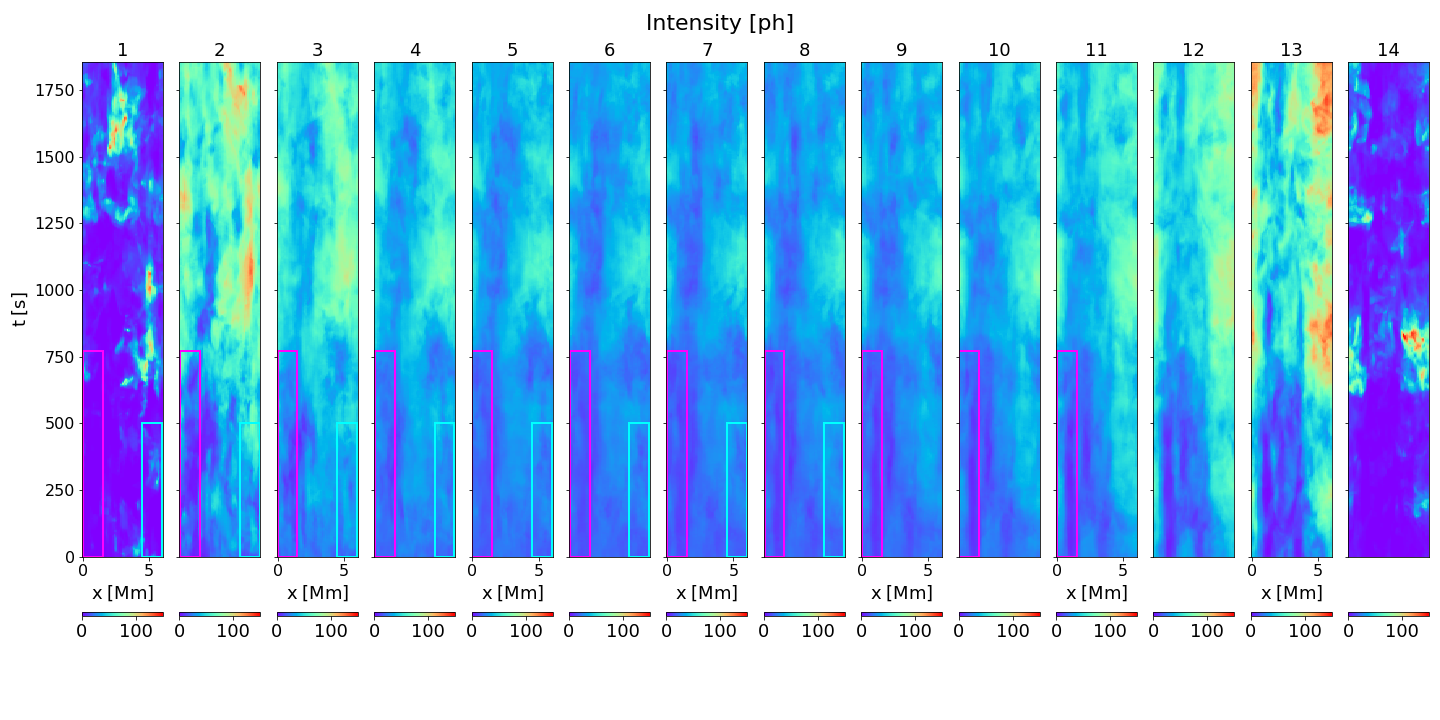}}
    \caption{Synthetic "sit and stare" observation for 14 of the 35 MUSE slits. From left to right, the figure shows the time-distance plots of the intensity for each slit. The location of the slits is marked by the dashed black lines in Fig. \ref{fig:raster_scan}. The pink and blue boxes mark the location of long-lived strong Doppler shifts.}
    \label{fig:emiss}
\end{figure*}

\begin{figure*}
	\resizebox{\hsize}{!}{\includegraphics{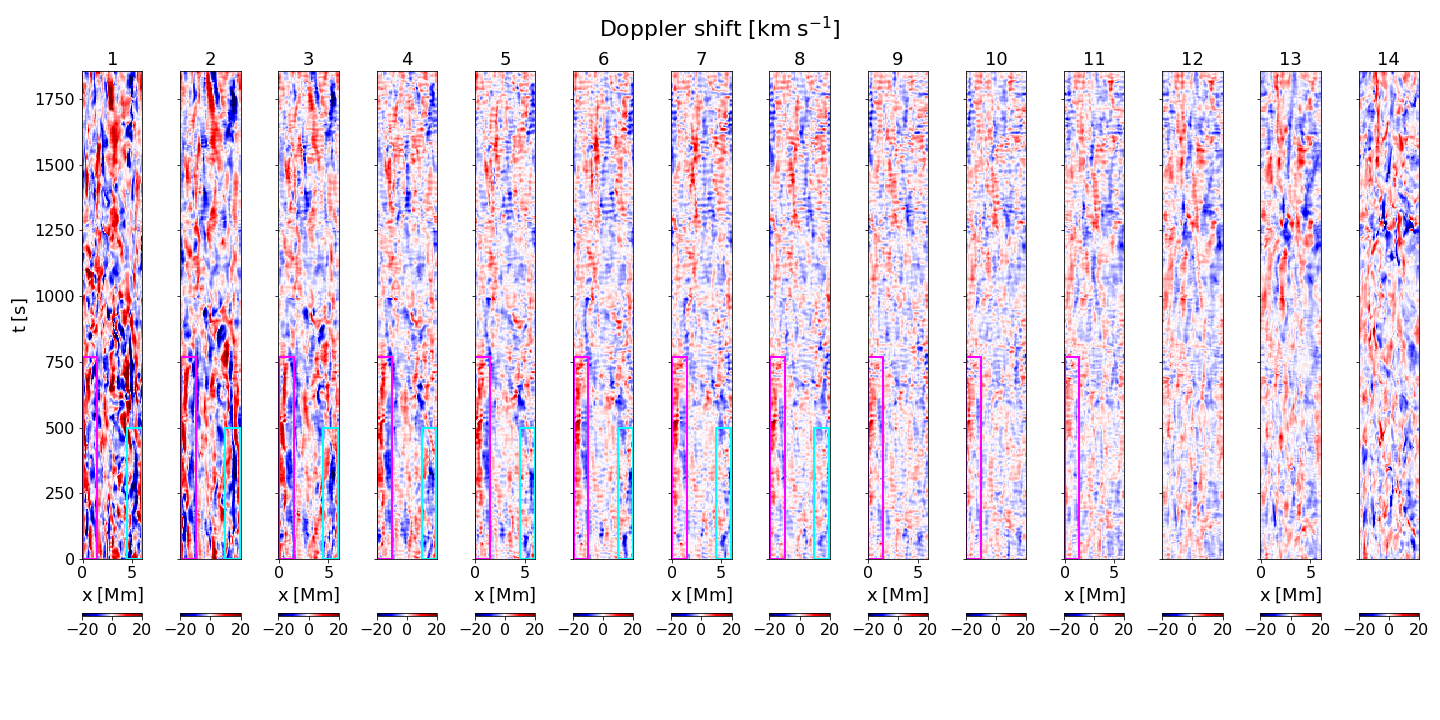}}
    \caption{Synthetic "sit and stare" observation for 14 of the 35 MUSE slits. From left to right, the figure shows time-distance plots of the Doppler shift at each slit. The location of the slits is marked by the dashed black lines in Fig. \ref{fig:raster_scan}. The pink and blue boxes mark the location of long-lived strong Doppler shifts.}
    \label{fig:doppler}
\end{figure*}

\begin{figure*}
	\resizebox{\hsize}{!}{\includegraphics{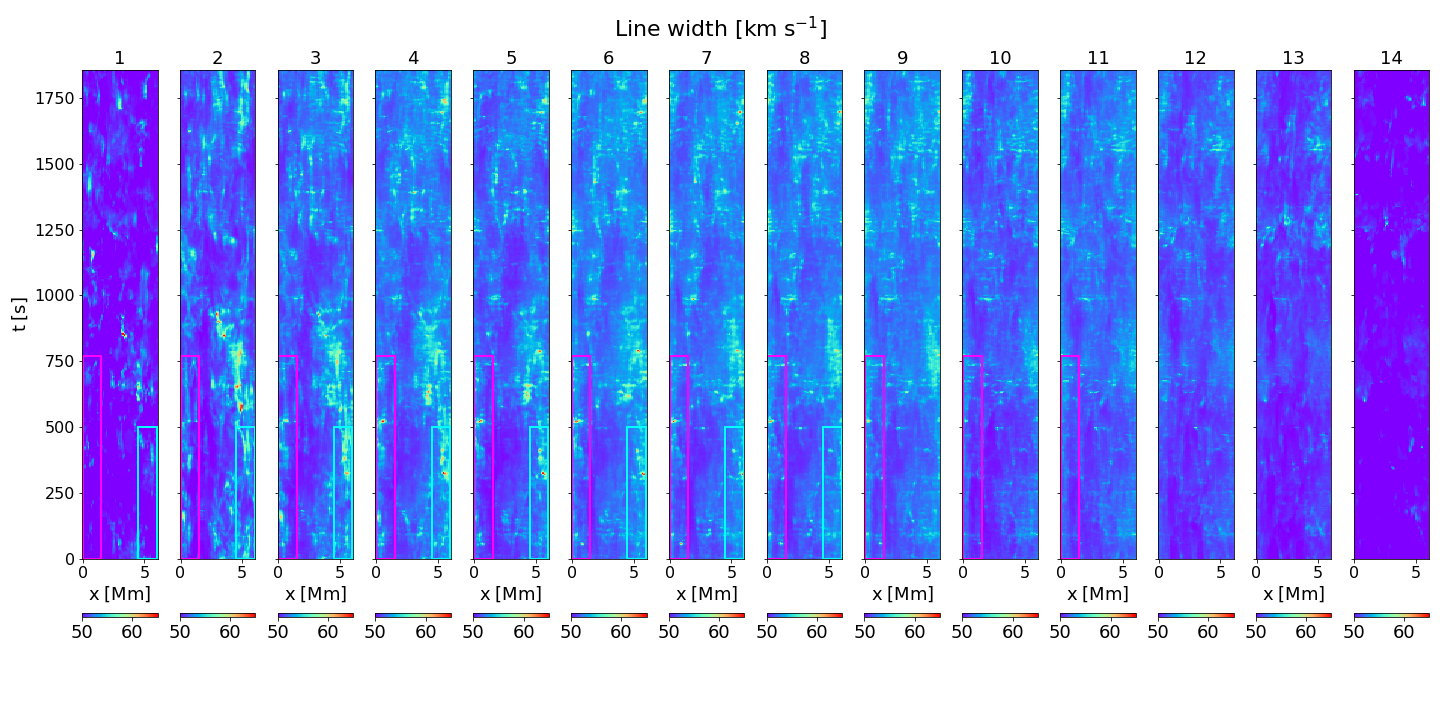}}
    \caption{Synthetic "sit and stare" observation for 14 of the 35 MUSE slits. From left to right, the figure shows the time-distance plots of the line width at each slit. The location of the slits is marked by the dashed black lines in Fig. \ref{fig:raster_scan}. The pink and blue boxes mark the location of long-lived strong Doppler shifts.}
    \label{fig:linewidth}
\end{figure*}

\begin{figure}
	\includegraphics[width=\columnwidth]{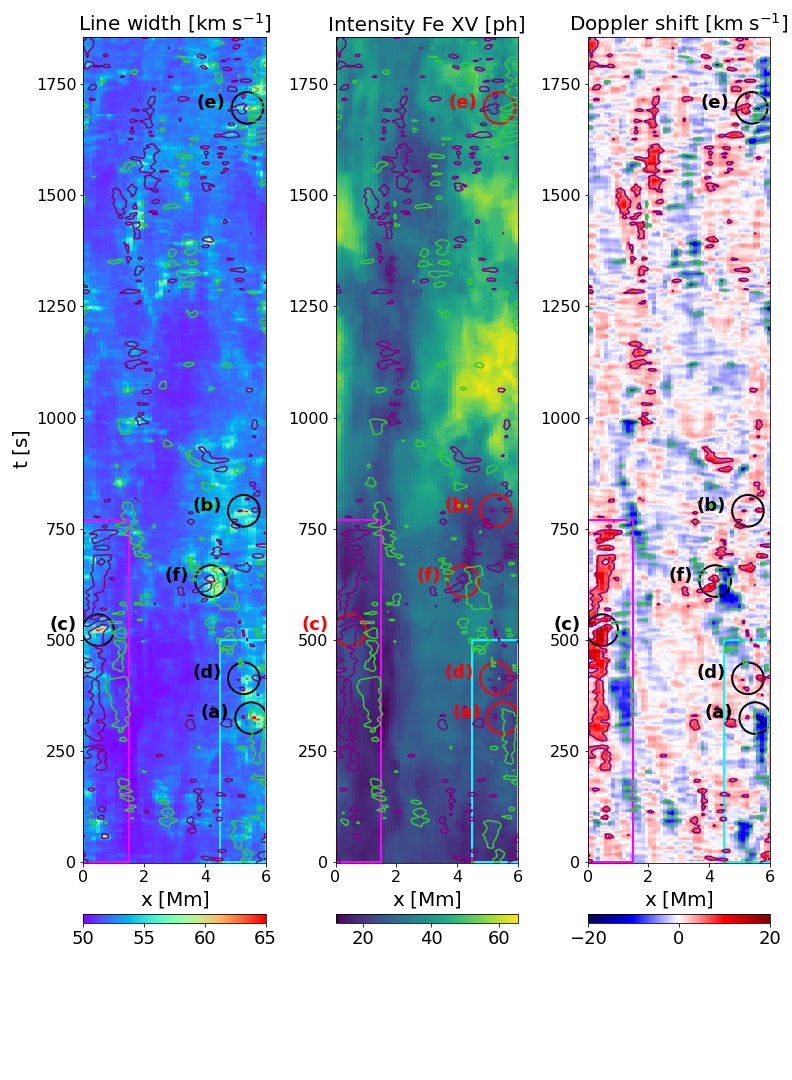}
    \caption{Synthetic "sit- and stare" observation for slit 5. From left to right: Line width for the \fexv line, Doppler shift and Intensity. The locations of the six strongest heating events are marked with black and red circles. The pink and blue boxes mark the location of long-lived strong Doppler shifts. The green and purple contours outline areas in the time-distance plots with Doppler shifts above the threshold of $5\; \rm{km\; s^{-1}}$.}
    \label{fig:events}
\end{figure}

We calculate an artificial \fexv\  MUSE raster scan and sit-and-stare observations of the simulated flux tube.
We identify an elongated structure with high line width in the raster scan at $x\approx 5.5\; \rm{Mm}$ (light blue rectangle in Fig. \ref{fig:raster_scan}). 
At the location of this structure, the synthetic emission displays elongated parallel blue- and redshifted features with adjacent blue- and redshift, indicating the possible presence of twisted structures. 
There is a second blue- and redshifted feature at x=0-1.5 Mm (pink rectangle), but only in the first case is the feature cospatial with a region of strong line broadening. There is no clear relation between Doppler shift, line width and the intensity.\\
We can also identify several persistent adjacent red- and blueshifted features in the time-distance diagram for the Doppler shift. 
Time-distance diagrams for a sit-and-stare observation for all 14 slits are shown in Fig. \ref{fig:emiss} - \ref{fig:linewidth}.
The largest and most long-lived feature, located at x=0-1.5 Mm, is visible from slit 1 through 11 (pink rectangles), indicating a spatial extent along the loop axis of about 40 Mm, and lasts for several minutes. The feature appears for different time ranges in different slits. While it is present for about 800 s in slit 2, it appears fainter in slit 11 and appears for a shorter time of only 200 s. This feature also appears in Fig. \ref{fig:raster_scan} (pink box), extending roughly 30 Mm along the loop. The second feature appearing in the raster scan (light blue box) is also visible in the time series, but from slit 5 on only the blueshifted component is discernible (blue rectangles).\\
Similar to the raster scan, regions with increased line broadening tend to be cospatial with strong Doppler shifts, although similar to the raster scan, the intensity shows no obvious correlation with Doppler shift or line width.\\
As an example, we
choose the timeseries for slit 5 (Fig. \ref{fig:linewidth}) at an axial distance of $s=16.545\; \rm{Mm}$ from the photosphere for event detection. Slit 5 is marked in orange in Fig. \ref{fig:raster_scan}. 
This part of the loop is in the corona and would be visible in an on-disk observation. 
We identify 36 events using the methods outlined in section \ref{sect:datanalysis}. The six strongest events are marked in Fig. \ref{fig:events} with circles. Of the total of 36 events, 34 show patterns of alternating red- and blueshifts within a range of 500 km around the spike in the line width. Since Doppler shift fluctuations are present nearly everywhere in the domain, we could also consider a threshold on the amplitude of the Doppler shift as an additional selection criteria. Using a threshold of $5\; \rm{km\; s^{-1}}$ to only take into account stronger flows reduces the number of identified events to 13 (out of 36).
Only in four cases are the peaks in line width associated with local peaks in intensity. Some events are associated with local minima in the intensity, and four cases are even close to or coincide with a global minimum in the intensity along the slit. Out of the six strongest peaks in line width, two occur at the location of a local intensity peak. Five out of the six events are associated with a sign change in the Doppler shift under consideration of the $5\; \rm{km\; s^{-1}}$ threshold.\\

\subsection{Selected Events}

\begin{figure*}
    \resizebox{\hsize}{!}{\includegraphics{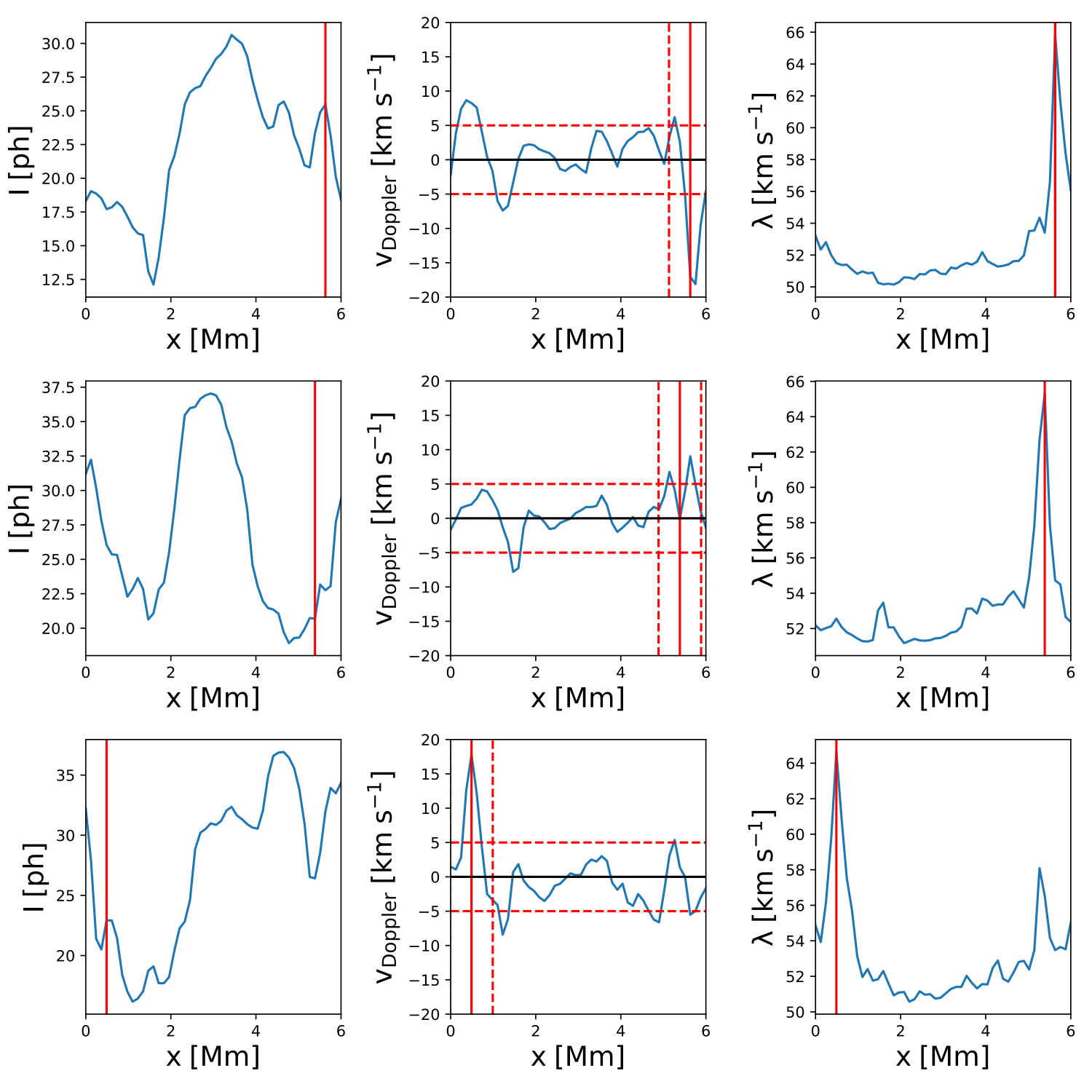}}
    \caption{From left to right: Cuts through intensity, Doppler shift and line width at the position of slit 5 for the \fexv\ line for events (a)-(c) (top to bottom). The vertical solid line marks the position of the highest peak in the line width. The dashed vertical lines in panel 2 mark a distance of 0.5 Mm from the peak in line width. The horizontal dashed lines show the threshold on the magnitude of the Doppler shift of $5\; \rm{km\; s^{-1}}$.}
    \label{fig:cut_sl1}
\end{figure*}

\begin{figure*}
    \resizebox{\hsize}{!}{\includegraphics{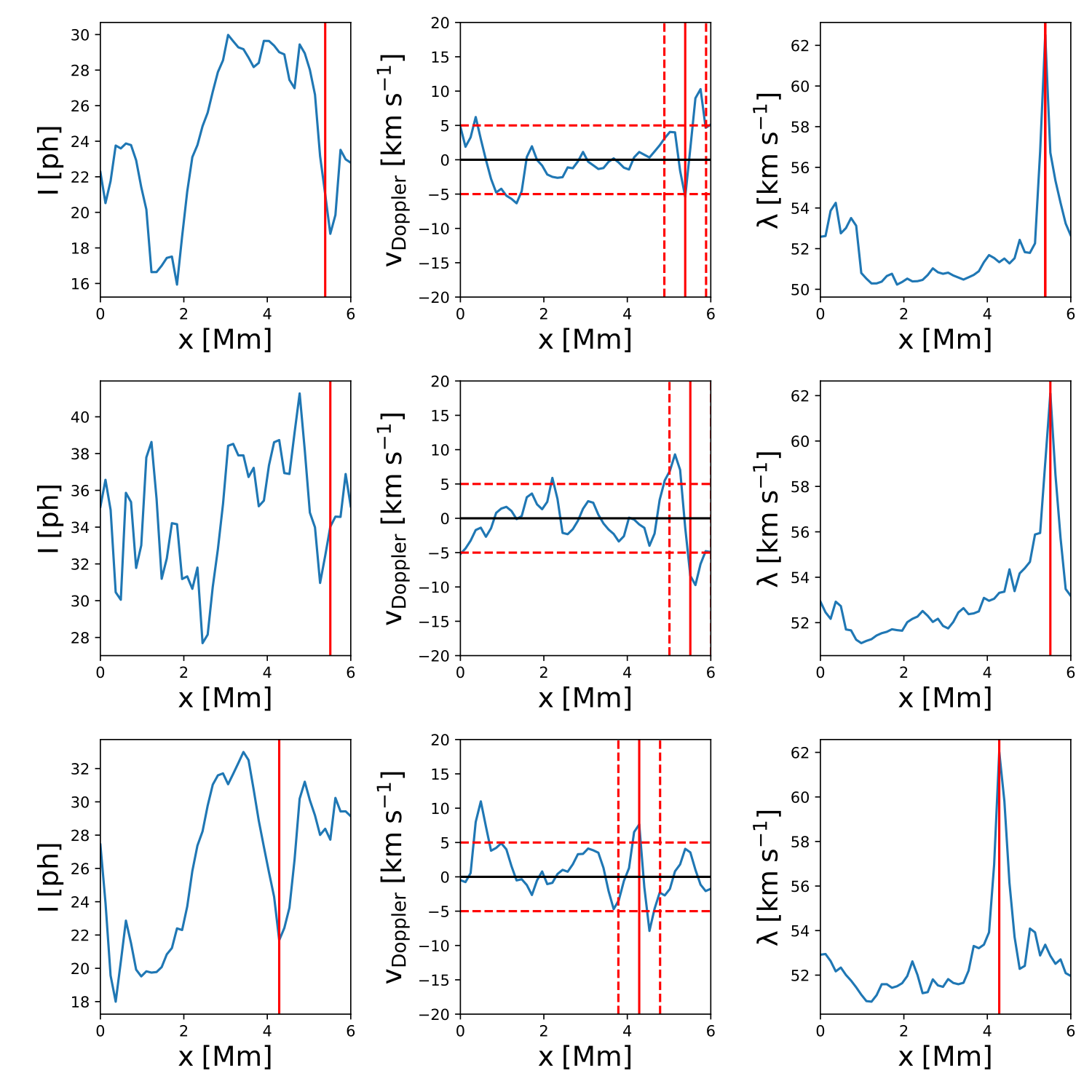}}
    \caption{From left to right: Cuts through intensity, Doppler shift and line width at the position of slit 5 for the \fexv\ line for events (d)-(f) (top to bottom). The vertical solid line marks the position of the highest peak in the line width. The dashed vertical lines in panel 2 mark a distance of 0.5 Mm from the peak in line width. The horizontal dashed lines show the threshold on the magnitude of the Doppler shift of $5\; \rm{km\; s^{-1}}$.}
    \label{fig:cut_sl2}
\end{figure*}

\begin{figure*}
	\resizebox{\hsize}{!}{\includegraphics{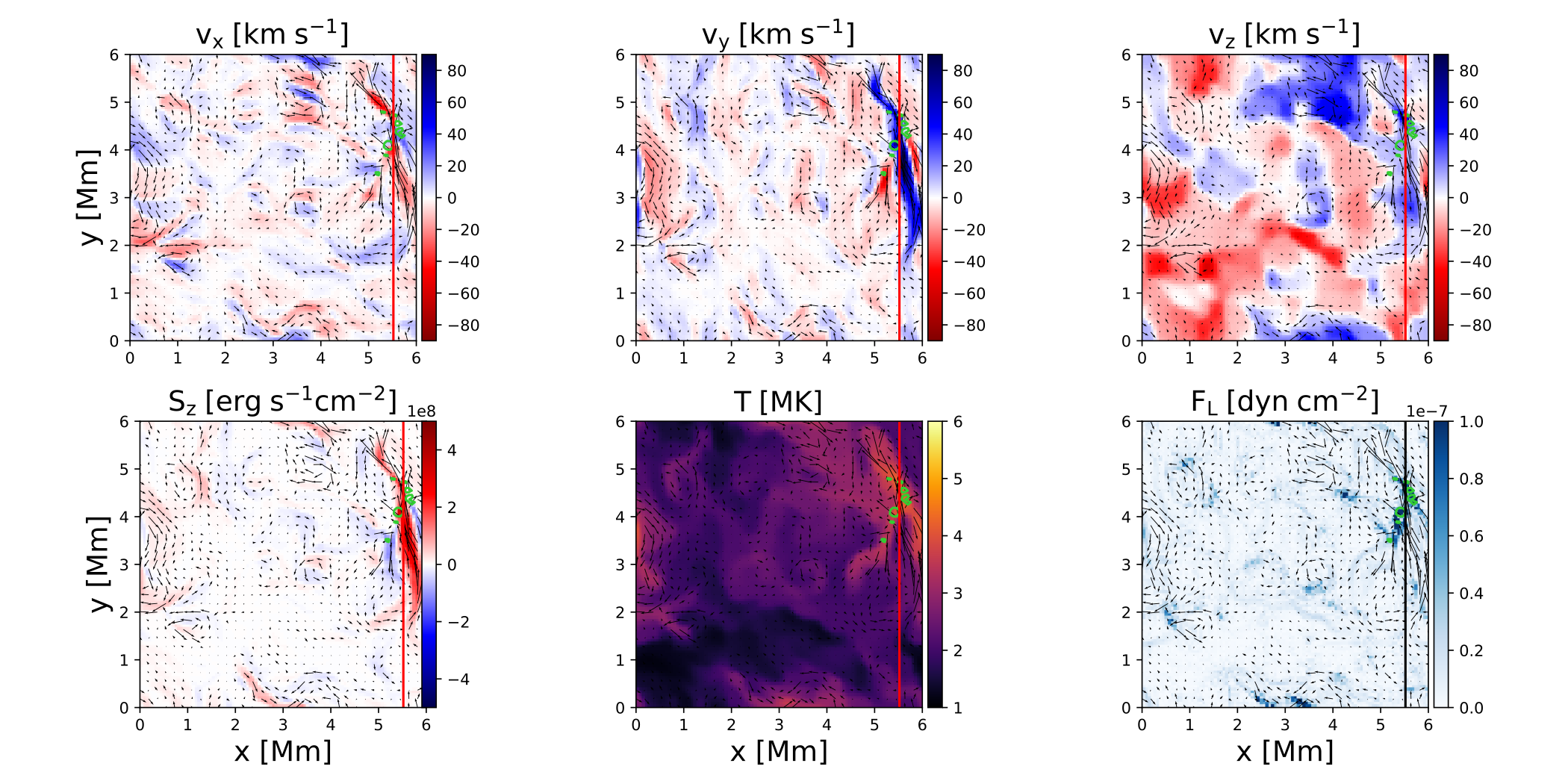}}
    \caption{Cuts perpendicular to the loop axis at the location of slit 5 for event (a). Top row: x, y and z component of the velocity. Bottom row: Axial Poynting flux, temperature, and horizontally directed Lorentz force. The arrows illustrate the velocity field. The solid red line indicates the position of the peak in line width for this event. The green contour outlines regions with a heating rate of $log_{10}(Q_{tot})>-1$.}
    \label{fig:cuts_perp1}
\end{figure*}

\begin{figure*}
	\resizebox{\hsize}{!}{\includegraphics{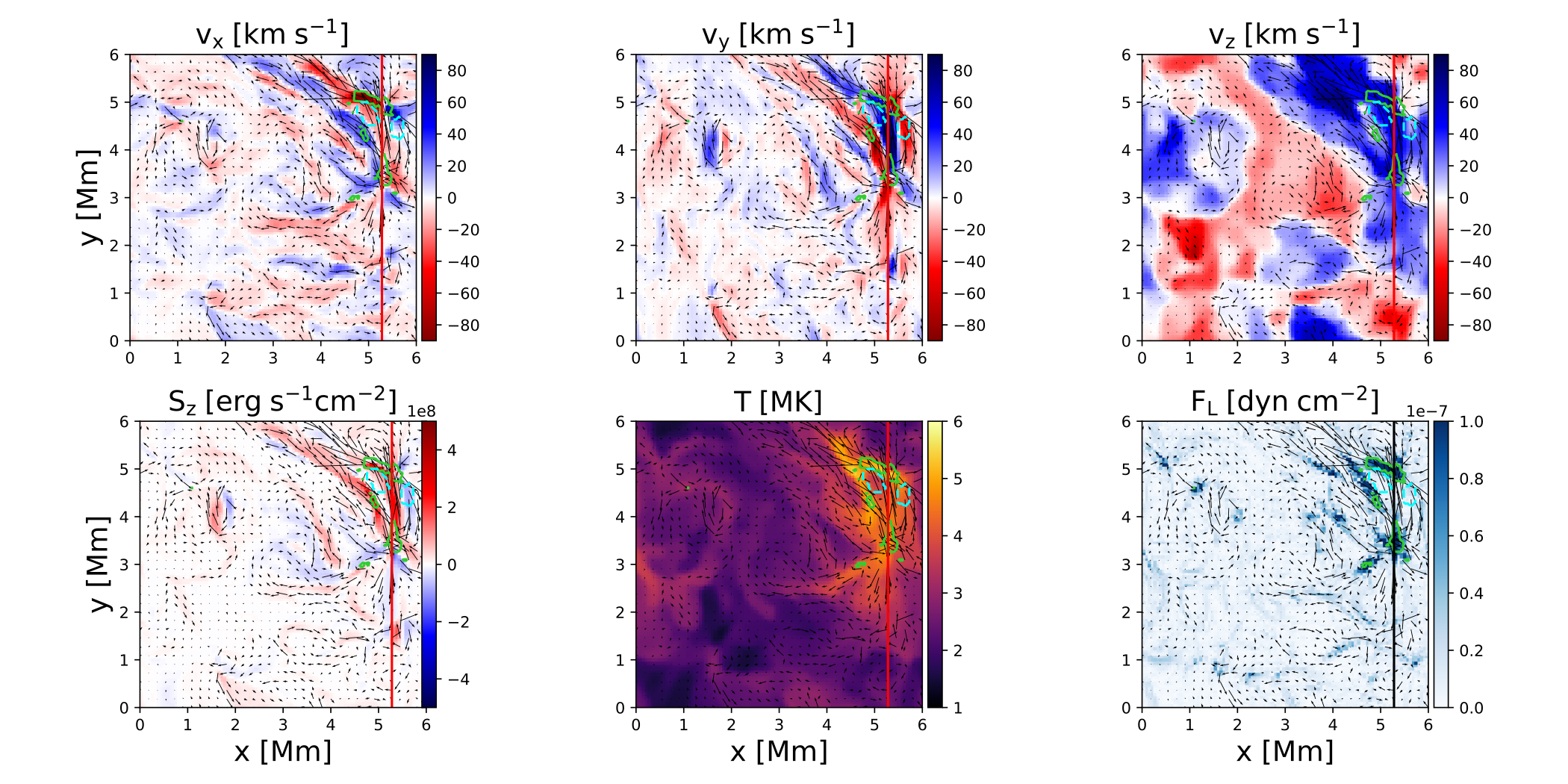}}
    \caption{Cuts perpendicular to the loop axis at the location of slit 5 for event (b). Top row: x, y and z component of the velocity. Bottom row: Axial Poynting flux, temperature, and horizontally directed Lorentz force. The arrows illustrate the velocity field. The arrows illustrate the velocity field. The solid red line indicates the position of the peak in line width for this event. The green contour outlines regions with a heating rate of $log_{10}(Q_{tot})>-1$. The dashed contour marks the region from which field lines are traced in Fig. \ref{fig:chrom_connect}}.
    \label{fig:cuts_perp2}
\end{figure*}

\begin{figure*}
	\resizebox{\hsize}{!}{\includegraphics{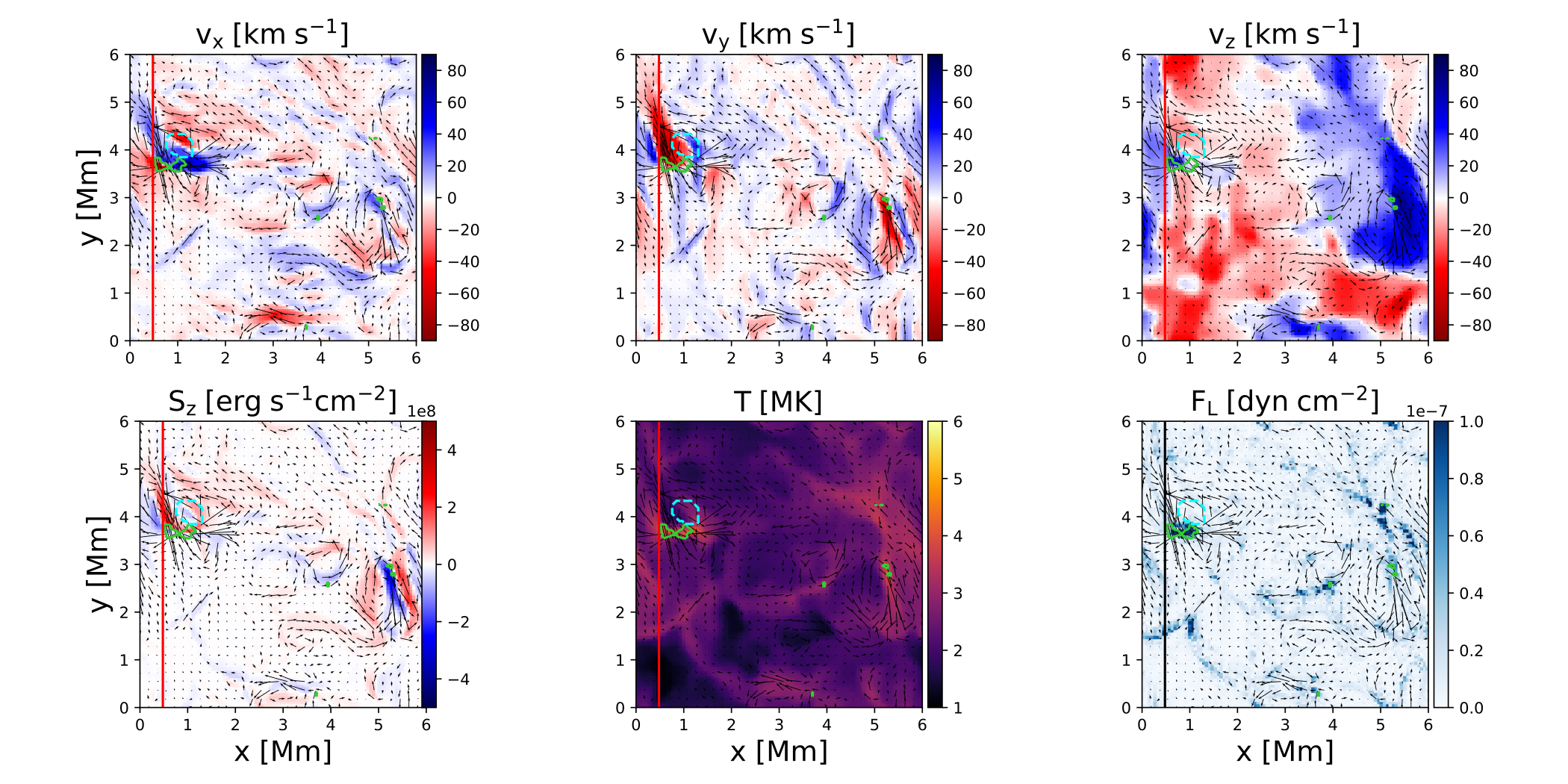}}
    \caption{Cuts perpendicular to the loop axis at the location of slit 5 for event (c). Top row: x, y and z component of the velocity. Bottom row: Axial Poynting flux, temperature, and horizontally directed Lorentz force. The arrows illustrate the velocity field. The solid red line indicates the position of the peak in line width for this event. The green contour outlines regions with a heating rate of $log_{10}(Q_{tot})>-1$. The dashed contour marks the region from which field lines are traced in Fig. \ref{fig:chrom_connect}.}
    \label{fig:cuts_perp3}
\end{figure*}

We choose the six strongest events for closer examination. Fig. \ref{fig:raster_scan} shows a MUSE raster scan centered around the strongest event (a). 
An elongated structure with high line widths as well as blue and red shifted emission is present along almost the full length of the loop at x=4.5 Mm and y=0-40 Mm. The alternating blue- and redshift indicates a potential twisted structure.
Intensity, Doppler shift and line width are shown at the time of peak broadening for events (a) to (c) in Fig.  \ref{fig:cut_sl1} and for events (d)-(f) in Fig. \ref{fig:cut_sl2}.  The peak in line width coincides with a peak in Doppler shift in case (a) and (c). For event (a) and (c), the peak in line width is at the location of a local maximum in the intensity, while for event (b) it is located close to the global minimum of the intensity.\\
The enhancement in the line width can be caused by different kinds of flow patterns.
Line broadening can arise from the cumulative effect of fluctuations in the plasma flow along the direction of the LOS or from small-scale flows below the instrument resolution. Since the resolution in this initial study is relatively low with 60 km, the effect is expected to arise mainly from the LOS-integration. Line broadening can also arise from increased plasma temperature during heating events. We found that the total line width for the \fexv\ line integrated over the coronal part of the loop is always above the thermal width for a viewing window covering the coronal part of the loop for a similar setup \citep{2024MNRAS.tmp..911B}. We checked for the three strongest line broadening events in our time series that this is also the case for individual heating events by computing the total line width from our simulation cubes assuming the temperature is constant everywhere and equal to the line formation temperature. We find a comparable peak in the line width in the same location and conclude that the enhancement in line width is not primarily caused by an increased plasma temperature and we can neglect the effect of thermal broadening. The MUSE response functions also contain the effect of instrumental broadening, but this is uniform in space.
\\
We find that in most cases the strongest line broadening occurs near sign changes in the Doppler shift.
 Torsional motions,  bidirectional jets such as nanojets and shear flows could all lead to parallel red- and blueshifted features adjacent to each other and to high line widths. For a bidirectional jet this would be the case if the LOS is along the direction of the jet axis so that both the component moving towards the observer and away from the observer are detected.\\
 In order to determine whether the detected events are related to small-scale swirls or other types of flow, we have a closer look at events (a)-(c).
Cuts perpendicular to the loop axis are shown in Figs. \ref{fig:cuts_perp1}-\ref{fig:cuts_perp3} for various quantities.

\begin{figure}
    \resizebox{\hsize}{!}{\includegraphics{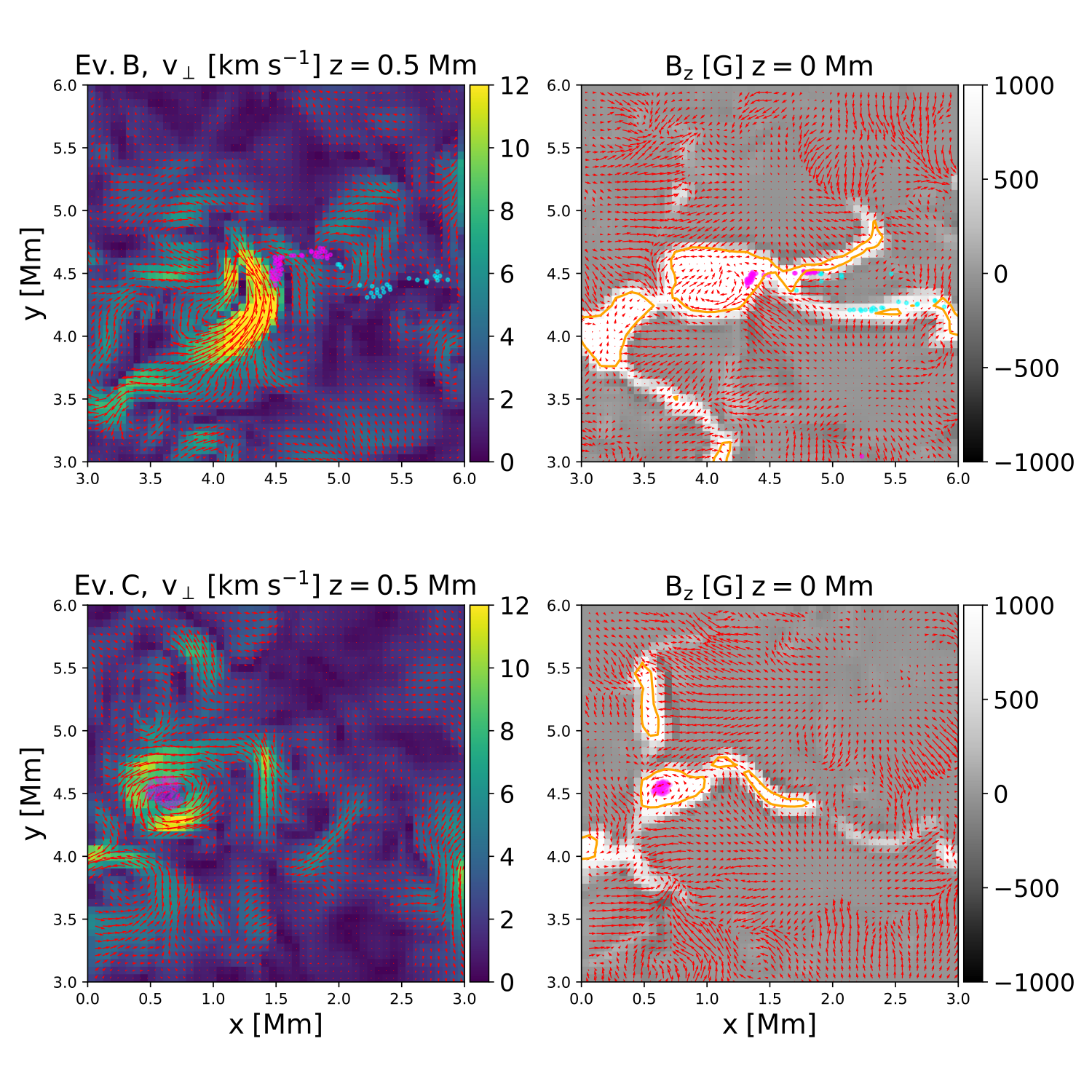}}
    \caption{Top row: Transverse velocity at a height of 0.5 Mm and vertical magnetic field at the photosphere for event (b). Bottom row: Same as the top row for event (c). The pink and blue markers mark the intersection of the magnetic field lines with a slice at height 0.5 Mm (left column) and the photosphere (right column). The orange contours outline kilogauss magnetic field concentrations.}
    \label{fig:chrom_connect}
\end{figure}

For event (a), a strong shear flow is present at the location of the increased line width and Doppler shift. The shear flow leads to the strong shift seen in the time-distance diagram at $x\sim 5.5\; \rm{Mm}$. The line of sight also crosses a strong heating event marked by green contours of the volumetric heating rate in Fig. \ref{fig:cuts_perp1}. Temperature and vertical velocity are enhanced at the location of the shear flow. Additionally, the component of the Lorentz force perpendicular to the loop axis is enhanced, driving outflows with speeds up to $80\; \rm{km\; s^{-1}}$. \\
Event (b) consists of a superposition of swirling/shearing flows and jet-like outflows (see Fig. \ref{fig:cuts_perp2}). The swirling component is strongly elongated. Temperature and vertical velocity are enhanced and two strong heating events are present along the LOS.\\
A flow consisting of two counter-rotating vortices is present in event (c), also showing some jet-like features driven by an enhanced Lorentz force (see Fig. \ref{fig:cuts_perp3}). Event (c) is also associated with a strong heating event at the interface of the vortices.
Both up- and downflows are present at the location of the heating event.\\
Upwards directed Poynting flux is enhanced for all three events.
We rarely find isolated swirl events, superpositions of swirls are common.\\
To test whether events (b) and (c) have a chromospheric counterpart, we traced field lines from a cut through the events at the height of slit 5 down to the chromosphere and photosphere. The starting points for the magnetic field lines were selected to lie in regions with enhanced swirling strength (here we set a threshold of $0.002\; \rm{rad\; s^{-1}}$, which corresponds to a rotation period of less than 50 min). In order to detect only the larger structures, we smoothen the velocity field with a Gaussian with an FWHM of 500 km before computing the swirling strength.
We then calculate the intersection of the field lines with a slice at a height of 0.5 Mm and with the photosphere.
The result is illustrated in Fig. \ref{fig:chrom_connect}. While for event (c), the magnetic field lines are clearly rooted in a chromospheric region and a kilogauss magnetic field concentration exhibiting swirling motions, the case is less clear for event (b).
Event (b) consists of a superposition of several swirls,  and the magnetic field lines are rooted in a larger, more complex footpoint. We trace field lines from two identified swirls. One of them appears to be connected to the edge of a photospheric swirl. In the chromosphere, it is connected to a flow that exhibits partial swirling motions, but is weaker than a nearby large-scale swirl.
We conclude that while some coronal swirls are magnetically connected to a swirling structure in the chromosphere, others show no clear connection and might form in the corona itself due to an energy cascade to small scales or local wave excitation.\\
Strong Doppler shifts are present several hundred seconds before event (c). In the supplementary movie, a superposition of several slow-moving rotating flows is present along the LOS in the region x=0-2 Mm and y=0-6 Mm. 
At the time of event (c), the heating event leads to strong outflows in opposite directions along the LOS. While the swirls present before event (c) are mostly aligned and show the same sense of rotation, the strong outflows occurring at the time of event (c) lead instead to a broadening of the emission line profile.

\subsection{Hot and Cool Plasma}

\begin{figure*}
	\resizebox{\hsize}{!}{\includegraphics{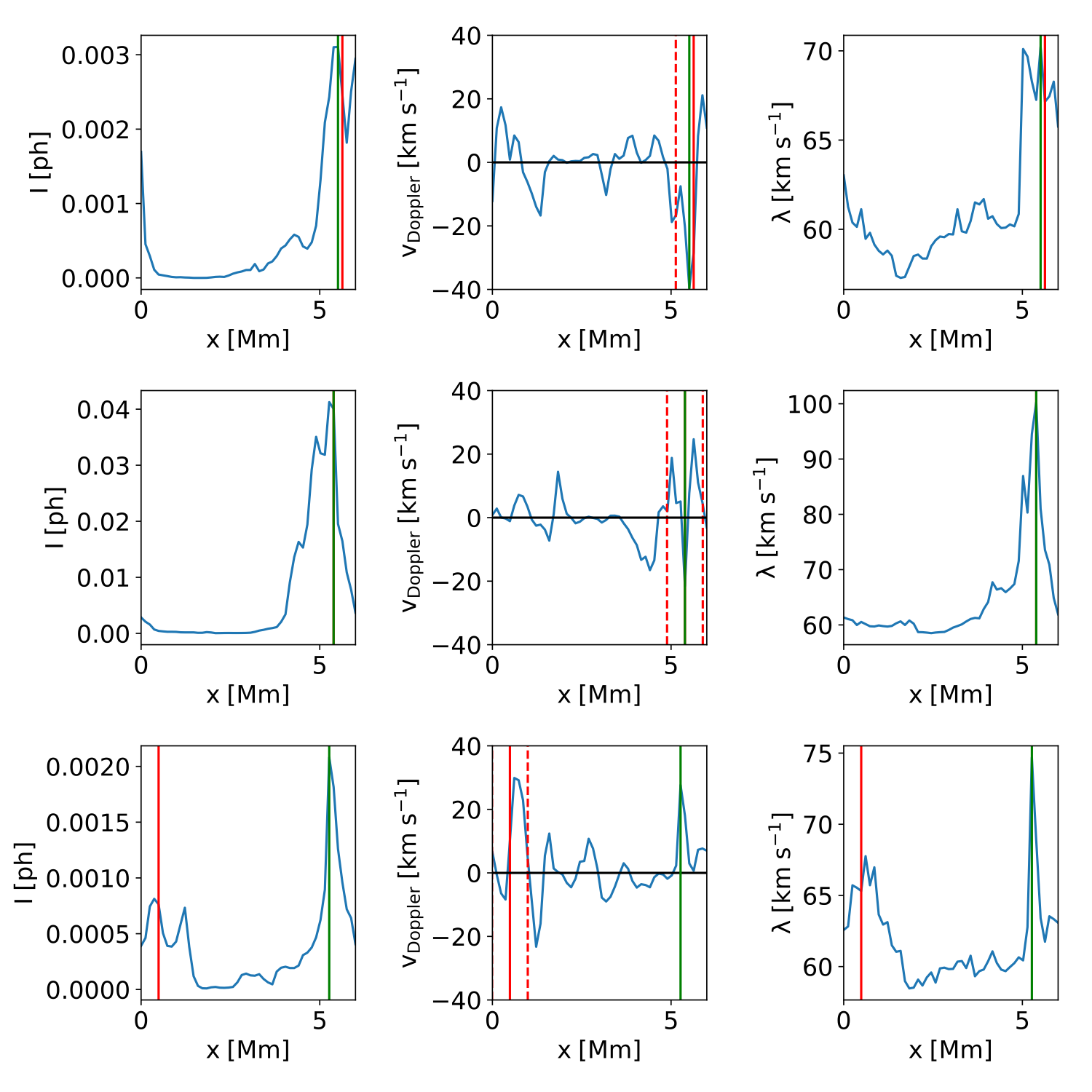}}
    \caption{From left to right: Cuts through intensity, Doppler shift and line width at the position of slit 5 for the \fexix\ line for events (a)-(c) (top to bottom). The vertical red line marks the position of the highest peak in the line width for the \fexv\ line. The vertical green line shows the location of the largest line width along the slit for the \fexix\ line. Only the green line is shown whee the red and green line overlap. The dashed red lines in panel 2 mark a distance of 0.5 Mm from the peak in line width.}
    \label{fig:cut_sl_FeXIX}
\end{figure*}

Broadening of coronal emission lines is most likely caused by small-scale motions that could be associated with heating. Previous studies find some correlation of nonthermal line broadening and intensity \citep{1984ApJ...281..870D, 1998ApJ...505..957C,2016ApJ...827...99T}, which we do not find in our simulation for the \fexv\ emission. 
There is no clear correlation between large line width, Doppler shift, and brightenings for the \fexv\ line. Most peaks in line width and Doppler shift are not associated with a peak in intensity.
The lack of correlation between Doppler shift, line broadening and intensity could be due to the fast-moving plasma responsible for the increases in Doppler shift and line broadening exceeding the peak formation temperature of the \fexv\ ion.
The presence of very hot plasma at temperatures of several million Kelvin is a signature of heating by nanoflares  \citep[e.g.,][]{1994ApJ...422..381C,2009ApJ...698..756R,2009ApJ...693L.131S,2012ApJ...750L..10T,2023hxga.book..134T} that locally heat loop strands to flare temperatures. We find plasma exceeding temperatures of 2.5 MK along the LOS for all of the six events under closer consideration. For event (b) and (f), the plasma temperature even exceeds 5 and 6 MK, respectively. Plasma of this temperature should be bright in the \fexix\ line, since the peak formation temperature of this ion is in the range of $\log_{10}(T) [K]=[7,7.1]$.\\ While most line broadening events are not associated with brightenings in \fexv\, we should find a correlation for hotter lines such as \fexix.
We have synthesized spectra for the \fexix\ line.  Intensity, Doppler shift and line width at the position of slit 5 are shown in Fig. \ref{fig:cut_sl_FeXIX} for events (a) to (c). In contrast to the \fexv\ emission, large line widths are associated with brightenings in the \fexix\ emission. We found a strong correlation between line width and \fexix\ intensity for the events (a) to (c). Due to the small filling factor and consequentially low photon count rates, emission in the \fexix\ channel would not be measurable for this loop.\\ 
Since the loop also contains plasma below the peak formation temperature of \fexv, we checked the emission in the \feix\ line with a peak formation temperature of 0.8 MK. Due to the sensitivity to cooler plasma, the emission in this line is concentrated mostly near the footpoints for our model, with negligible emission in the coronal part of the loop. 
 
\subsection{Exposure time}

In order to determine Doppler shift and line width with an accuracy of $5\; \rm{km\; s^{-1}}$, roughly 100 detected photons are needed for the \feix\ line and 150 for the \fexv\ line (see Fig. 6 in \citealt{2022ApJ...926...52D}).

\begin{figure}
	\includegraphics[width=\columnwidth]{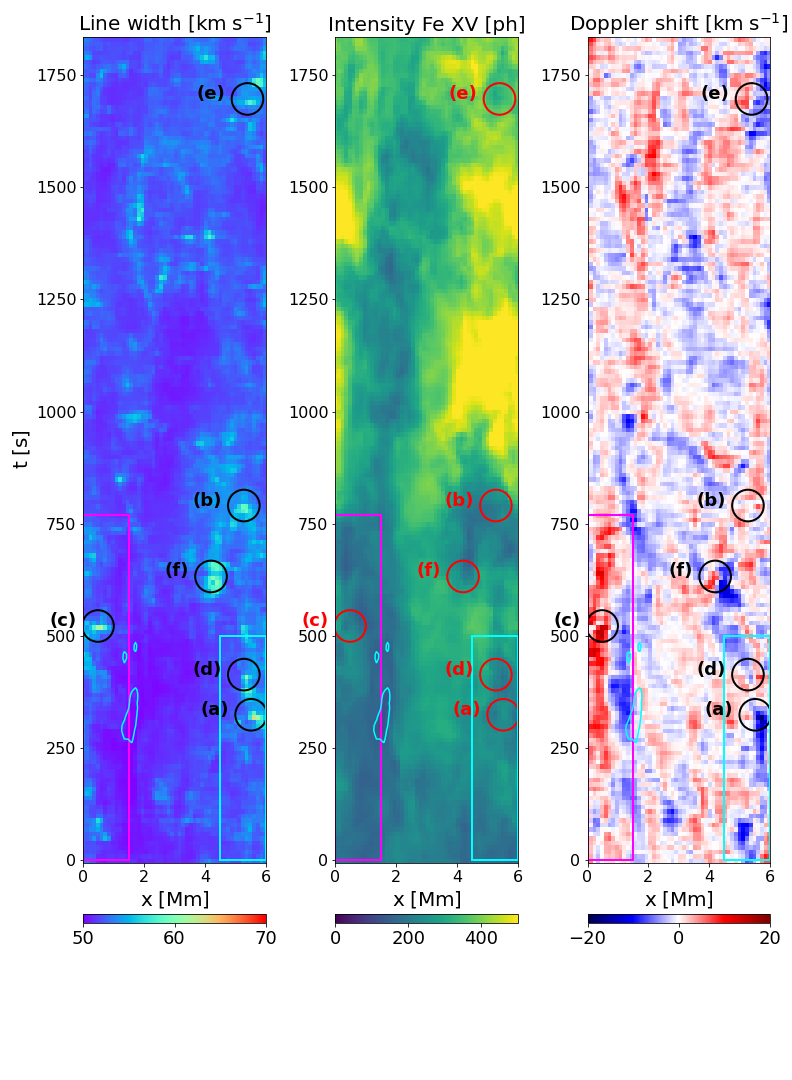}
    \caption{Time-integrated time-distance images for the \fexv\ line at slit 5. Left to right: Intensity, Doppler shift, and line width. We assume an exposure time of 10 seconds. The pale blue contours mark regions where the photon count is below a threshold of 150 photons. Detected events are marked with circles.}
    \label{fig:time_int}
\end{figure}

We test different exposure times to check how long we need to integrate in time to obtain a sufficient amount of photons to determine Doppler shift and line width with the desired accuracy.
Since heating events occur on short timescales, it is important to not expose for too long, so that the events are still  discernible as separate heating events.
For an exposure time of 5 s, the majority of detected events lie outside of areas where enough photons are detected. An integration time of at least 10 s is needed to achieve the desired accuracy. The time-distance diagrams for the intensity, Doppler shift, and line width are shown in Fig. \ref{fig:time_int} for an integration time of 10 s. A sufficient number of photons is detected for most of the loop area.
Using the same event detection method as before, we recover the events (a)-(f) for the time-integrated data. 
 \\

\subsection{Time evolution and propagating features}

\begin{figure*}
	\resizebox{\hsize}{!}{\includegraphics{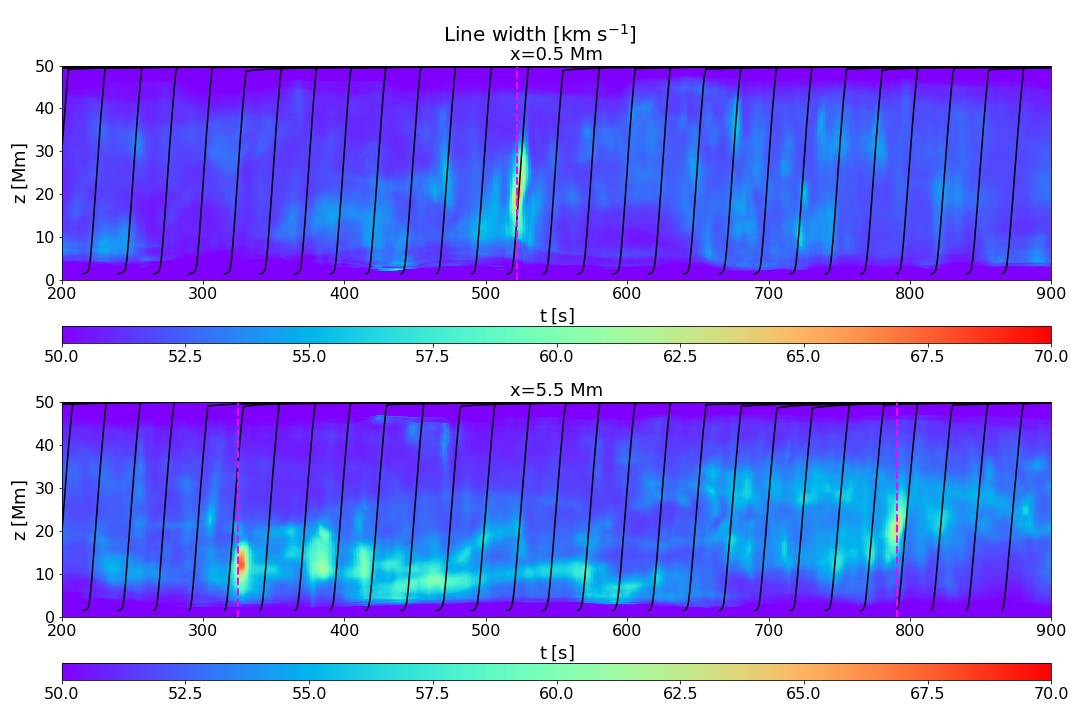}}
    \caption{Time-distance image for the line width of the \fexv\ line for two slits placed parallel to the loop axis at positions x=0.5 Mm and x=5.5 Mm. The dashed red lines mark the time of the three strongest line broadening events. Event (a) and (c) are occurring near x=5.5 Mm, while event (b) occurs near x=0.5 Mm. The times of the events are marked with dashed vertical
    lines. The trajectories of test particles moving with the Alfv\'{e}n speed are overplotted.}
    \label{fig:vert_slit}
\end{figure*}

\begin{figure*}
        \resizebox{\hsize}{!}{\includegraphics{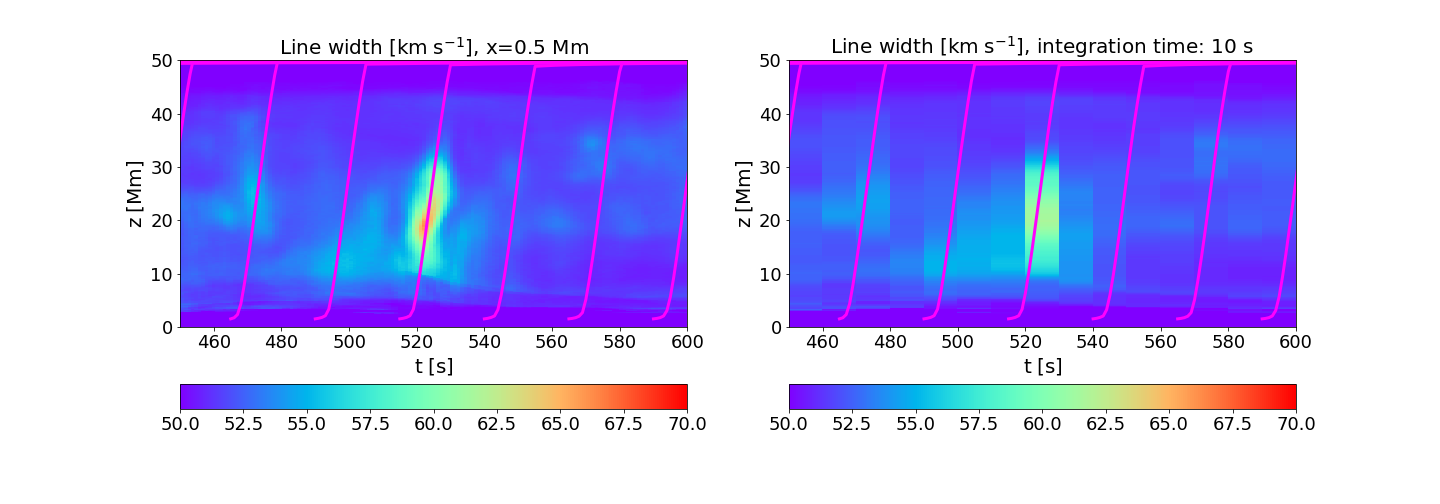}}
    \caption{Time-distance image for the line width of the \fexv\ line at a slit placed parallel to the loop axis at position x=0.5 Mm. Left panel: Zoom-in of Fig. \ref{fig:vert_slit}. Right panel: Time-distance image for the time-integrated spectra. The data has been integrated over $\sim 10\; s$. The overplotted magenta curves represent tracks for test particles moving with the time-dependent Alfv\'{e}n velocity averaged along the LOS.}
    \label{fig:vert_slit_zoom}
\end{figure*}

While in the lower atmosphere individual swirls are discernible, with height the flow field becomes increasingly complex.
Instead of being associated with a single isolated twisted structure, the region with high line width and Doppler shift  at x=4-6 Mm and y=3-6 Mm shows many emerging and disappearing rotating structures (see Figs. \ref{fig:cuts_perp1}-\ref{fig:cuts_perp3} and accompanying movie).
The distinction between swirls and waves is not clear and depends on timescales of photospheric motions and propagating disturbances in the corona. While in the low atmosphere a swirl might be a persistent structure, the Alfv\'{e}n speed increases steeply at the transition region and twists in the magnetic field are quickly propagated away. 
\\
We are interested in the question whether the swirls we see in our simulations are contiguous, persistent twisted structures, torsional oscillations, or propagating "Alfv\'{e}n pulses" as suggested by \citet{2021A&A...649A.121B}, and whether it would be possible to detect their propagation if they are indeed features moving upwards from the photosphere.\\
Most line broadening events and strong Doppler shifts can be seen in several slits. The broadening appears at slightly different times for slits at different positions along the loop.\\
To check whether this delay is associated with propagating features,  we plot a time-distance diagram assuming that the instrument slits are oriented parallel to the loop axis.  \\
A time-distance diagram for the line width for two loop-aligned slits located at x=0.5 and x=5.5 is shown in Fig. \ref{fig:vert_slit}. The slits are placed to cover the location of events (a)-(c).\\
The time-distance diagrams shows several elongated structures. Since heating events usually occur as extended structures along the loop axis, these signals could be signatures of spatially extended intermittent heating events or propagating features. 
In order to check if these signals are consistent with perturbations propagating with the Alfv\'en speed, we calculate the trajectories of test particles moving with the local, time-dependent Alfv\'{e}n speed as $\Delta z = (v_{A}+v_{z})\Delta t$. These trajectories are overplotted in Fig. \ref{fig:vert_slit} as black lines and their slope is indeed roughly compatible with the slope of the line broadening events (a)-(c).
Depending on whether or not the Doppler shift undergoes a sign change, we can distinguish between a torsional oscillation and a pulse.
Instead of the characteristic pattern expected for a torsional oscillation, we find unidirectional Doppler shifts lasting for several 100 seconds. \\
A practical obstacle to detect signals of propagating disturbances lies in the necessary exposure time to achieve the photon count required to measure Doppler shifts and line width with sufficient accuracy.
For an integration time of five seconds the majority of detected events occur in regions that do not show a sufficient photon count.
This poses a problem for the detection of propagating disturbances due to the high Alfv\'{e}n speed and therefore short travel times in the corona.
While fewer maxima for the line width are detected in the time-integrated data due to the loss of fine structure, the strongest events can still be detected even after 10 s exposure time.
The detection of signatures of propagation, however, is impeded for an exposure time of 10 s. As illustrated in Fig. \ref{fig:vert_slit_zoom} for event (c), we would still detect the enhancement in line broadening, but the structure in the time-distance diagram does not appear inclined anymore and it is therefore not possible to distinguish between a propagating feature or an event that occurs simultaneously over a large height range.
With Alfv\'{e}n speeds of the order of thousands of kilometers and a loop length of 50 Mm, the travel time is of the order of 10 s. An event would need to have a travel time of more than 20 s along the flux tube in order to be detected. Likewise, we can only detect oscillations with periods above 20 s.
The coronal field strength of 60 G we choose for our simulation, however, is quite high and corresponds to active region loops. For active region loops, measured coronal electron densities are $\log_{10}(n)\; = 8.9-9.8 \,\rm{[cm^{-3}]}$ and field strengths in the range of 60-150 G \citep{2021ApJ...915L..24B}, while being on the order of a few G in other regions \citep{2020Sci...369..694Y} and several 100 G in flare loops \citep{2019ApJ...874..126K}, leading to large variations in typical Alfv\'{e}n speeds.

\subsection{Discussion and Outlook}

Swirls cause observable signatures in Doppler shift and line width. They could be observed as elongated structures showing adjacent red and blueshifts in the Doppler velocity, enhanced line widths, or a combination of both. 
The exact nature of swirls is not yet solved, they might be stationary rotating magnetic flux tubes or torsional Alfv\'{e}nic waves \citep{2021A&A...649A.121B}. Some of the structures we find in our simulation appear to be propagating along the loop axis roughly at the Alfv\'{e}n speed. Due to the high cadence that MUSE allows for, it may be possible to follow perturbations in time, but the low intensity of these structures and the resulting long required exposure times might prohibit this.\\
Similar signatures to the swirl signatures we find, however, could also be produced by a jet or shear flow depending on the observing angle. Additional uncertainty is introduced by the overlap of emission from different structures along the LOS due to the optically thin nature of the coronal emission. The Doppler shift is nonzero and fluctuating  in almost every part of the domain and the Doppler shift signal alone is not enough to detect presence of swirl.
A few events, however, stand out due to the spatial and temporal coherence of enhanced Doppler shifts that hint at the presence of a persistent flow structure instead of a superposition of randomly directed flows.
Most line broadening events are located near a region of strong Doppler shift or at its edges. The presence of both enhanced Doppler shifts and enhanced line widths suggests both flows on large scales as well as heating and associated small-scale flows.\\
Distinguishing between different kinds of events depends on information about the 3D structure of the observed plasma flows. A 3D reconstruction would require observations from multiple different aspect angles. This could be achieved with a combination of MUSE and SPICE \citep{2020A&A...642A..14S} or IRIS \citep{2014SoPh..289.2733D} for cooler lines. The spectral resolution of SPICE, however, is not sufficient to accurately determine the line widths in non-flaring coronal loops. MUSE could also be combined with an imager in order to determine plasma velocities by tracking motions of bright structures (see \citealt{2024A&A...681L..11E}), but this method is not applicable for purely torsional Alfv\'{e}n waves that do not exhibit transverse motions and do not lead to brightenings. MUSE observations would have to be supplemented by additional observations such as chromospheric observations in the same region to shine more light on the nature of the events. An observation of a chromospheric swirl would make the presence of a similar structure in the corona more likely. 
Furthermore, it has been suggested that swirls could result from a merger of photospheric magnetic concentrations \citep{2022A&A...665A.118F,2023ApJ...949....8K}. These structures have a  small diameter of hundreds of kilometers, but could be observed with new instruments such as DKIST and EUVST, which will make it possible to trace plasma from transition region to coronal temperatures \citep{2019SPIE11118E..07S,2022ApJ...926...52D}. This combination of instruments should make it possible to identify coherent structures spanning different atmospheric layers.
\\
The numerical model itself has several limitations. Here we use a relatively low resolution of 60 km due to the lower computational cost of obtaining a run with a high temporal cadence.
With higher numerical resolution, the velocity field is expected to become increasingly complex and more small-scale vortices will be resolved, leading to more overlapping structures along the LOS.\\
 A problem with detecting the very hot plasma component that is bright in \fexix\ arises from its small emission measure and the short time range over which it is present \citep{2014LRSP...11....4R}. \\
 
 The plasma spends a much longer time in the subsequent cooling phase than in the initial very hot phase. Doppler shift and line width are highest in the first 800 seconds, while the \fexv\ emission reaches its highest value after that time, indicating that this increase could be due to cooling nanoflare-heated plasma. Optically thin coronal emission has a strong dependency on the density. The timescale  of the evaporation of plasma from the denser chromosphere is longer than the typical duration of heating events. This explains why the loop reaches its maximum intensity after the heating activity indicated by high Doppler shifts and line broadening subsides.\\
 The correlation between Doppler shifts, line width and intensity for the \fexix\ line confirms that the lack of correlation for the \fexv\ line is due to the plasma temperature far exceeding the peak formation temperature of \fexv\ ion. The photon count rate from our model, however, is too small to be observed due to the low filling factor of the hot plasma. Nevertheless, \fexix\ observations could be relevant for very hot active region loops.
 Since the \feix\ emission is very weak in the coronal part of the loop, 
 the \fexv\ line is therefore the most suitable line available to observe the simulated warm loop. Even for this line, however, the photon count rates are low compared to count rates expected for active regions (see e.g., for comparison \citealt{2020ApJ...888....3D,2022ApJ...926...52D}). Count rates of a few tens of photons per second increase the exposure time needed to accurately measure Doppler shifts and line widths to at least ten seconds. The low count rates compared to active region simulations could be caused by the small line-of-sight integration for this model of just 6 Mm. In active region simulations, plasma with temperatures around 1.5-2 MK is present along the LOS for about 20 Mm (see Fig. 9 in \citep{2017ApJ...834...10R}). Our box would cover only a small part of that loop system. Since the loop is multithermal, the filling factor of plasma at the right temperature to be captured in a given passband is low. Additionally, coronal densities are expected to be underestimated in the model since the transition region is underresolved \citep{Bradshaw_2013}.\\
 %Due to the optically thin nature of coronal emission, the signatures of multiple structures along the LOS overlap. 
The stretched loop setup sacrifices realism for a lower computational cost for resolving the loop interior.
The model uses a uniform vertical magnetic field as initial condition. Therefore, the field strength does not vary strongly and the field is close to vertical already at chromospheric heights. The field strength in the corona relative to the field strength at the photosphere is therefore likely overestimated. With 50 Mm, the loop length corresponds to a typical active region loop length. The plasma temperature of around two MK also corresponds to active region loops, but we do not include sunspots in the simulation box.
Numerical experiments have to be conducted using a realistic curved loop setup in order to determine the influence of the magnetic topology on swirl properties and observables.\\
Currently, we assume quite idealized conditions when conducting the forward modelling. The response function we use to compute the synthetic emission incorporates only the main emission line. We do not take into account other lines or the overlapping of spectra from different slits on the detector. This contribution from additional lines should be taken into account in order to make realistic predictions about observations, although its effect is expected to be largely negligible in most spectra \citep{2020ApJ...888....3D}, and, where that is not the case, it can be estimated by applying a spectral disambiguation code \citep{2019ApJ...882...13C, 2020ApJ...888....3D}.
%contamination %has to be taken into account in order to make realistic predictions about observations. For now, we assume that the spectral disambiguation code \citep{2020ApJ...888....3D} gets rid of this effect.\\ 
We also did not add noise to the line profiles, which would be present in actual observations. This study therefore represents an idealized situation and needs to be expanded upon by including more instrumental effects.

\section{Conclusions}

The aim of this paper is to investigate whether MUSE could detect coronal swirls.

In synthetic emission derived from our numerical model, we find multiple instances of  line broadening events cospatial with parallel features with adjacent strong red- and blueshifts. We find signatures of the propagation of some of these features to higher atmospheric layers in time-distance diagrams. These events are usually associated with shear flows or a superposition of many small scale swirls. Despite swirls having been linked to coronal heating, they are not always associated with a brightening in the respective emission line. This is due to the multithermal nature of the plasma.
It is a challenge for observations to obtain a high enough photon count to accurately measure line shifts and widths.
Longer exposure times needed for faint contributions from hot plasma complicate the detection of propagation signatures in regions with strong field and low density, leading to high Alfv\'{e}n propagation speeds. Stereoscopic observations would be needed to verify that an observed event is a swirl and not due to a jet or shear flow.
Despite these limitations, MUSE could potentially observe the limb counterpart of small-scale swirls or could detect their presence in coronal loops under favorable conditions, e.g., for brighter events.

%We estimate that with integration times of 10 s, the required photon count can be achieved.

\section*{Acknowledgements}
The authors would like to thank Juan Mart\'{i}nez-Sykora for
the MUSE response functions and python scripts for calculating spectral line profiles.\\ The research leading to these results has received funding from the UK Science and Technology Facilities Council (consolidated grant ST/W001195/1). The research leading to these results
has received funding from a Royal Society Wolfson Fellowship (RSWF/FT/180005). IDM received funding from the Research Council of Norway through its Centres of Excellence scheme, project number 262622.
PT was supported by contract 4105785828 (MUSE) to the Smithsonian Astrophysical Observatory, and by NASA grant 80NSSC20K1272.

%%%%%%%%%%%%%%%%%%%%%%%%%%%%%%%%%%%%%%%%%%%%%%%%%%
\section*{Data Availability}

Due to their size, the data from the numerical simulations and analysis presented
in this paper are available from the corresponding author upon
 request.



\bibliographystyle{mnras}
\bibliography{main} % if your bibtex file is called example.bib

%\begin{thebibliography}{99}
%\end{thebibliography}

%%%%%%%%%%%%%%%%%%%%%%%%%%%%%%%%%%%%%%%%%%%%%%%%%%

%%%%%%%%%%%%%%%%% APPENDICES %%%%%%%%%%%%%%%%%%%%%

\appendix



%%%%%%%%%%%%%%%%%%%%%%%%%%%%%%%%%%%%%%%%%%%%%%%%%%

% Don't change these lines
\bsp	% typesetting comment
\label{lastpage}
\end{document}